\documentclass[aps,prb,twocolumn,showpacs]{revtex4}
\usepackage{tabularx}
\usepackage{graphicx}
\usepackage{graphicx,subfigure}
\usepackage{dcolumn}
\usepackage{bm}
\usepackage{hyperref}
\hypersetup{
    colorlinks=true,
    linkcolor=blue,
    filecolor=magenta,
    citecolor = blue,      
    urlcolor=cyan,
}
\usepackage[all]{hypcap}
\usepackage{soul}

\begin{document}

\preprint{AIP/123-QED}

\title{Body-centered phase of shock loaded Cu}

\author{Anupam Neogi}
\email{anupamneogi@atdc.iitkgp.ernet.in} \affiliation{
Advanced Technology Development Centre, Indian Institute of Technology Kharagpur, Kharagpur- 721302, India
}%
\author{Nilanjan Mitra}%
 \email{nilanjan@civil.iitkgp.ernet.in}
\affiliation{%
Department of Civil Engineering, Indian Institute of Technology Kharagpur, Kharagpur- 721302, India
}%

\date{\today}

\begin{abstract}
Single crystal Cu when shock loaded in certain direction for a certain narrow range of piston velocities undergoes a structural phase transition from face-centered cubic to different body-centered phases. Based on molecular dynamic simulations, the manuscript identifies new phases of Cu through detailed radial distribution function and x-ray diffraction analysis. Identification of the new phases of Cu should initiate a re-evaluation of phase diagram for Cu under high temperature and pressure conditions.  
\end{abstract}

\maketitle

\section{INTRODUCTION}
A solid-solid diffusionless phase transition can be induced by a change in temperature and/or by application of external pressure. Shock loading of materials presents a unique situation in which the materials are subjected to extreme temperatures and pressures. Thereby it is possible that under suitable shock loading situations a metallic material may undergo structural solid-solid phase transition. Till now many elements have been identified to undergo this type of solid-solid phase transformation under different loading situations. Elemental Cu is not on the list of the elements which have been reported to exhibit solid-solid phase transition. The implications of solid-solid phase transformation are wide spread and have been used for defining a number of physical phenomena such as hardening, pseudoelasticity, shape memory effects; thereby benefitting numerous industries ranging from defense (e.g. strengthened steel using Cu-Fe alloys for marine ship hulls), infrastructure (e.g. Earthquake resistant structural dampers developed using Cu-based shape memory alloys), electronics (epitaxy of graphene over Cu film) and so on to name a few.

Copper is one of the most common elements worldwide. Because of the importance and widespread usage of the material, this element has been studied extensively (both experimental [see series of publications by Meyers and his group\cite{meyers1, meyers2, meyers3}] and numerical [see series of publications by Bringa, Germann, Ravelo and their groups \cite{bringa1, bringa2, bringa3, bringa4, germann1, germann2, germann3}]) along with different alloys it forms. Typically elemental Cu at ambient temperature and pressure exist as a face-centered cubic (FCC) material. Much of the above mentioned studies focus on formation and evolution of plasticity mechanisms (dislocation, twinning and different sessile loop formations) in shock loaded Cu. Structural phase transitions of Cu under shock loading have not been focused in the above mentioned researches.

Structural phase transitions (from FCC to body-centered cubic (BCC)/ body-centered tetragonal (BCT) phase) of Cu has been observed for Cu films grown epitaxially on Pd$\{001\}$, Pt$\{001\}$, Fe$\{001\}$ and Ag$\{001\}$\cite{wang, wu, tian, li}. On the other hand, $ab$-$initio$ studies have demonstrated that BCC phase (a special type of BCT phase) of Cu is energetically unstable at ambient temperature and pressure under tetragonal deformation\cite{lu, kraft} but stable under triagonal deformations. The apparent anomalous stability of a few atomic layers of BCC phase of Cu (pseudomorphically developed by epitaxy), in contrast to theoretical studies, has been reasoned as a typical glide of (110) BCC atomic planes over each other which usually takes place in the bulk case is prevented by the stable substrate surface playing the role of epitaxial constraint\cite{sob}. However a thick BCC layer can not be stabilized by this epitaxy and it reverts back to ground state FCC or some other truly metastable phase. Jona and Marcus\cite{jona} demonstrated the existence of a BCT phase of Cu with c/a = 0.93 which is tetragonally stable by calculation of the epitaxial Bain path of Cu. However it has also been mentioned in the manuscript since this special BCT phase does not satisfy all the stability criterions imposed on elastic constants, it is unstable against other modes of shear deformation. Typically these $ab$-$initio$ studies have been carried out at ground state; there maybe a possibility of stable body-centered phase of Cu at higher temperatures and pressures which is yet to be explored.  

Infact at higher temperatures, BCC phase of Cu is commonly observed for Cu-based alloys (Hume-Rothery materials) which define the usage of these materials for shape memory applications. Neutron diffraction investigations on BCC phase of Cu-Zn-Al\cite{guenin}, Cu-Al-Ni\cite{hoshino}, Cu-Al-Pd \cite{nagasawa}, Cu-Al-Be\cite{manosa} have demonstrated that the whole TA2[110] phonon branch has very low frequencies and softens with temperature as transition is approached. Friedel\cite{friedel} demonstrated that even though BCC phase of Cu is energetically unstable at the ground state, it may be preferred by the system at high temperature due to its large entropy resulting from low-energy vibrational transverse modes. Infact structural transition may be only driven by the excess of vibrational entropy of the high-temperature phase\cite{morris}. For Cu based shape memory alloys, Planes et al.\cite{planes} have shown that low energy of the TA2[110] phonon vibrational mode provides the major contribution to the excess of entropy which stabilizes BCC phase at high temperature. It should be noted that electron contribution to the entropy change is observed\cite{planes} to have little influence in driving the phase transformation for Cu based alloys. Through a series of calorimetric and magnetic measurements it has been identified that harmonic vibration of lattice is the main reason for stability of BCC phase in Cu based alloys\cite{ortin}. Within the context of Landau theory, Planes et al.\cite{vives} provided an experimental evidence that coupling between homogeneous shear and short wavelength phonon is an essential mechanism to account for martensitic transformations of Cu based alloys. Given the stable existence of BCC/BCT phase of Cu via epitaxy and as well as in Cu-based alloys, it is quite conceivable that there may be regions within the phase diagram of Cu where it exists in some body-centered form.

Taylor and Dodson\cite{taylor} first mentioned the possibility of formation and propagation of martensitic phase transition in bulk Cu behind a shock front. However their simulation study based on using many-body potentials for Cu subjected to shock loads identified phase transition based on common neighbor analysis (CNA). It should be noted that CNA or even adaptive-CNA (analysis algorithms based on nearest neighbors as available in LAMMPS code or many other MD codes) may not give a comprehensive representation of structural phase transition, which has been discussed later in the manuscript. The only way to systematically and accurately identify a new phase of a material is through X-ray diffraction (XRD) analysis, even though Radial distribution functions (RDF) and structure factor analysis may provide some important clues. Hirth et al.\cite{hirth} postulated that shock loading of Cu along the $<$110$>$ direction will lead to formation of a body-centered orthorhombic (BCO) phase, however a proper identification and detection of this novel phase of Cu has not been systematically dealt within the paper. Levitas and Ravelo\cite{levitas} proposed methodology of ``virtual melting'' to describe the loss of coordination number of the crystals behind the wave for a shock loaded Cu in the $<$110$>$ and $<$111$>$ directions. Through Hugoniotstat based molecular dynamic simulations Bolesta and Fomin\cite{fomin} criticized the concept of ``virtual melting'' and demonstrated nucleation of solid-solid phase transition behind a shock wave for a shock loaded polycrystal Cu using structure factor analysis. It should also be noted in here that the temperature required for melting of Cu at atmospheric pressure is significantly different from that required with shock compression in which huge pressures are generated (for a detailed study one is referred to An et al.\cite{he}). Typically a polycrystal Cu has numerous grains of different sizes oriented in different directions and separated by grain boundaries. The effects of grain size, grain orientation of the polycrystals of Cu under shock loads of different intensities has been investigated with regards to structural phase transition using only adaptive CNA techniques by Sichani and Spearot\cite{spearot}. 

It should be noted that the material responses observed in polycrystalline materials may not be same as that observed in monocrystalline materials since a polycrystalline material is a summation of different orientations of the monocrystalline material. If a behaviour is observed in a polycrystalline material then it is necessary to identify which orientation of the monocrystalline material is responsible for that particular behaviour. Questions may also be raised about whether this phase transition in polycrystal Cu is an artifact of grain size, pattern or is induced by the presence of different grain boundaries. Infact within this context, the works of Levitas and Ravelo\cite{levitas} dealing with single crystals in specific orientations cannot be directly correlated to the work of the Bolesta and Fomin\cite{fomin}. It is necessary to understand the physics of phase transition, if any, for different orientations of the single crystal prior to comments being made for polycrystals. The current work presents the first quantitative work in literature focusing on the phenomena of structural transitions in monocrystalline Cu under shock loads of different intensities in different directions. The current work not only identifies the novel phase of Cu from RDF analysis but also from XRD analysis. 

One of the most widely used potential for Cu, the EAM-Mishin potential \cite{mishini} has been used for this simulation. In this regard, it should be mentioned that Mishin EAM potential for Cu has been widely tested against $ab$-$initio$ and tight-binding methods and is well accepted by the research community for simulations of Cu. The potential shows good match with experimental elastic constants, experimental phonon dispersion curves thereby demonstrating its global reliability. Apart from calculation of lattice properties, the potential has also been extensively tested for various structural energies and transformation paths. For further details regarding the potential one should refer to the classic paper by Mishin et al. \cite{mishini} where details of the potential are provided.   

\section{SIMULATION METHODOLOGIES}
Non-equilibrium molecular dynamics (NEMD) simulation using EAM potential (developed by Mishin et al.\cite{mishini}) has been carried out to investigate the shock response of perfect monocrystalline copper in [100], [110] and [111] orientations at various piston velocities ranging from 0.3 km/s (leading to a pressure of 11.74 GPa and temperature of 384.39 K) to 2.5 km/s (leading to a pressure of 147.21 GPa and temperature of 3898.82 K). Since it is known that MD simulation results depends upon the choice of force potential, it should be mentioned that EAM potential by Mishin et al. \cite{mishini} has been rigorously tested and validated under different loading situations using $ab$-$initio$ as well as experimental investigations. There exists numerous studies using this EAM-Mishin potential for shock studies of monocrystalline Cu in different directions\cite{Erhart, murphy} with obvious validations to experimentally determined LASL shock Hugoniots data\cite{lasl}. A $500\times100\times100$ supercell of copper, with orthogonal axes X, Y and Z oriented along $<$100$>$, $<$010$>$ and $<$001$>$ respectively is created; equilibration of that simulation box is done with a timestep size of 1 fs by applying isothermal isobaric, NPT ensemble integration scheme in conjunction with Nose-Hoover thermostating algorithm at ambient temperature and pressure. The dimension of the sample is $180.76\times 36.15\times 36.15$ nm, $143.59\times 30.85\times 30.85$ nm and $175.86\times 30.85\times 30.85$ nm with  $2\times10^7$, $8064000$ and $16128000$ number of atoms for the $<$100$>$, $<$110$>$ and $<$110$>$ orientation of shock loaded cases respectively. Periodic boundary condition has been set along all three orthogonal directions during equilibration simulations. Standard momentum mirror technique\cite{neogi} is applied to produce shock waves. During shock compaction simulations, periodic boundary is applied along the directions normal to the shock i.e. Y and Z; along shock direction the boundary is kept fixed. Microcanonical (NVE) ensemble is used during shock simulation to ensure energy conservation throughout the time integration of the atoms. These shock simulations are continued till the shock front reaches the rear surface of the specimen sample.    

Typically one of the limitations of the NEMD simulations is that it cannot be run for larger time scales given its computational expensiveness. There is a possibility that a new structural phase observed in NEMD simulations may relax back to some other phase. Thereby in order to observe if similar behaviour is observed at larger timescales, multiscale shock technique (MSST)\cite{reed} algorithm has also been utilized for all corresponding simulation runs with smaller representative cell ($50\times 50\times 50$ crystal lattice units). The computational cell in MSST algorithm (implemented within the LAMMPS framework) follows a Lagrangian point through the shock wave which is accomplished by time evolving equations of motion for the atoms, as well as volume of the computational cell to constrain the stress in the propagation direction to the Rayleigh line and the energy of the system to the Hugoniot energy condition. Computational cell size for MSST simulation has been so chosen such that the stress, density, and energy density do not vary appreciably along the length of the computational domain (a requirement for MSST in which the stress and energy of a molecular dynamic simulation is constrained to obey the momentum and energy Hugoniot relations such that the simulation proceeds through the same thermodynamic states as would occur in a steady shock). Well equilibriated MD box (at ambient temperature and pressure) with before-mentioned crystallographic orientation is taken as initial configuration and equation of motion of atoms are integrated upto 10 ns. More theoretical foundation and technical details can be found elsewhere\cite{reed, neogi2, neogi3} for MSST implementation. It should be pointed out that no difference in observations can be made between the NEMD and the MSST simulation runs upto 10 ns. 

\section{RESULTS AND DISCUSSION}
As Cu monocrystals are subjected to shock load of varying intensities and in different directions, different types of plasticity mechanisms originate. The molecular dynamic simulations carried out by the authors matches with other numerical and experimental investigations carried out in this domain and details of new observations by the authors with regards to identification of different plasticity mechanisms with different shock intensities and shock loading directions has been submitted for publication elsewhere. Typically for low shock intensities (such as a shock velocity of 0.4 km/s) stress relief and energy relaxation is observed through formation of $\frac{a}{6}<$112$>$ Shockley partials at highly dense (111) planes which eventually starts gliding by generating imperfect lattice and stacking faults (leading to a stacking sequence of hexagonal closed packed (HCP) lattice behind those partials. Orthogonal intersections of slip planes occur resulting in interaction of partials and formation of stair-rod sessile dislocations $\frac{a}{6}<$110$>$, stacking fault tetrahedron (SFT) and Lomer-Cottrell (LC) locks $\frac{a}{3}<$110$>$ depending on the orientation of the lattice. 
\begin{figure}
\centering
  \includegraphics[width=7cm]{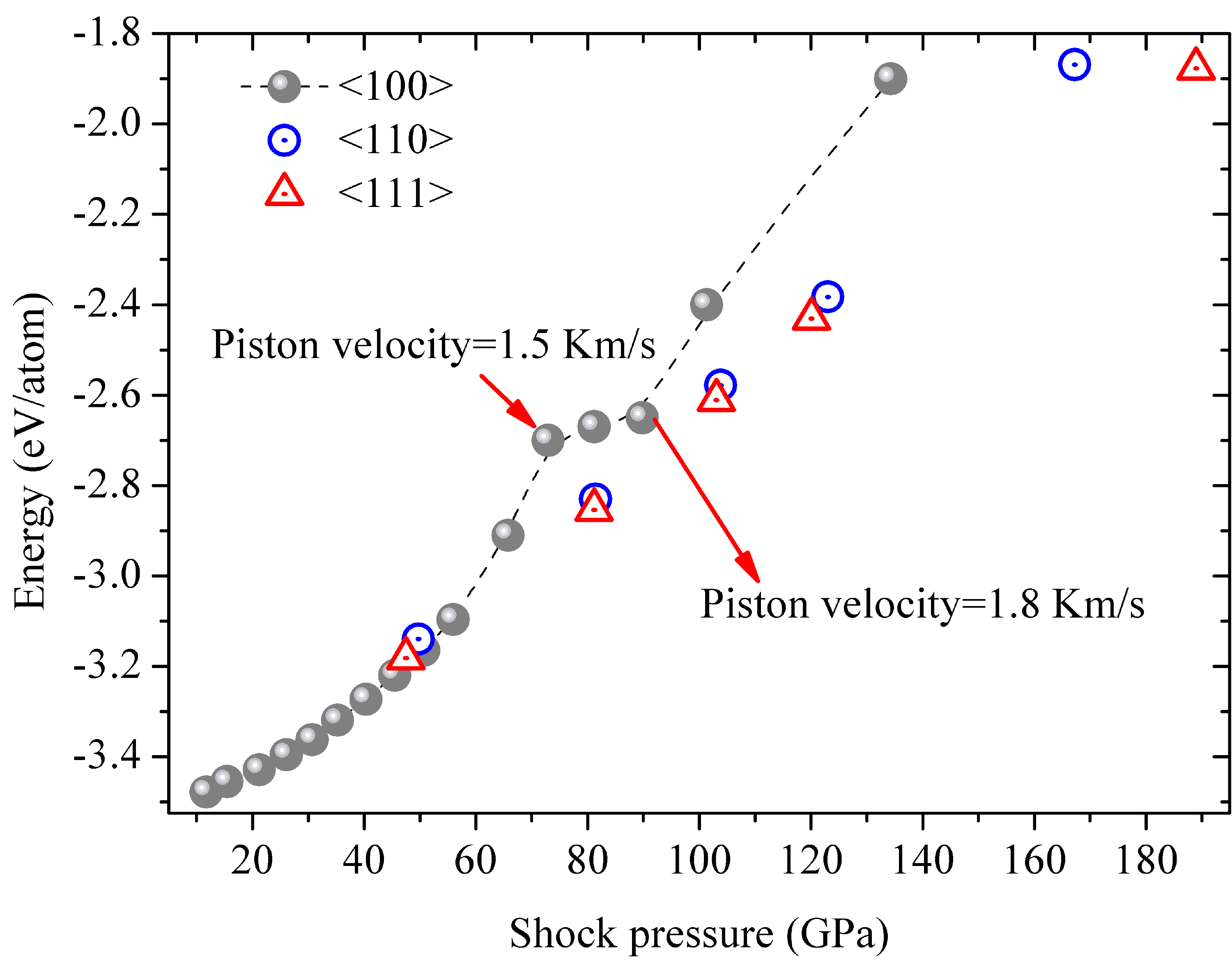}
  \caption{\label{pe} Shock pressure vs. potential energy (averaged over all atoms in deformed specimen) curve for shock loading along different directions $<$100$>$, $<$110$>$ and $<$111$>$ for different shock intensities ranging from 10-198 GPa. As written in the plot red color arrows represent observed 'plateau'-like region in between the impact velocity of 1.5 and 1.8 km/s for the shock direction of $<$100$>$.}
\end{figure}

At these low shock intensities, the adaptive-CNA technique\cite{Stukowski} (local lattice structure identification method suitable for multiphase system) for some atoms shows local (body-centered cubic) BCC and disordered structure with respect to initial FCC crystal symmetry (for a significant distance behind the shock front). No such signatures of body-centered crystal structure are obtained from RDF analysis (as discussed later in the manuscript) at these low intensity shock speeds. Typically energy at these high temperatures is distributed into kinetic and potential energy components. This eventually leads to fluctuation of neighbor distances, as a result of which the algorithm based on counting of nearest neighbors show inappropriate orders and thereby may incorrectly reveal that FCC lattice has lost its neighbors to show a BCC structure. Thereby, even though common neighbor based analyses might provide an initial rough estimate with regards to the phase of the material, it may not be reliable especially at high temperature and pressure situations.
\begin{figure*}
\centering
  \includegraphics[width=16.5cm]{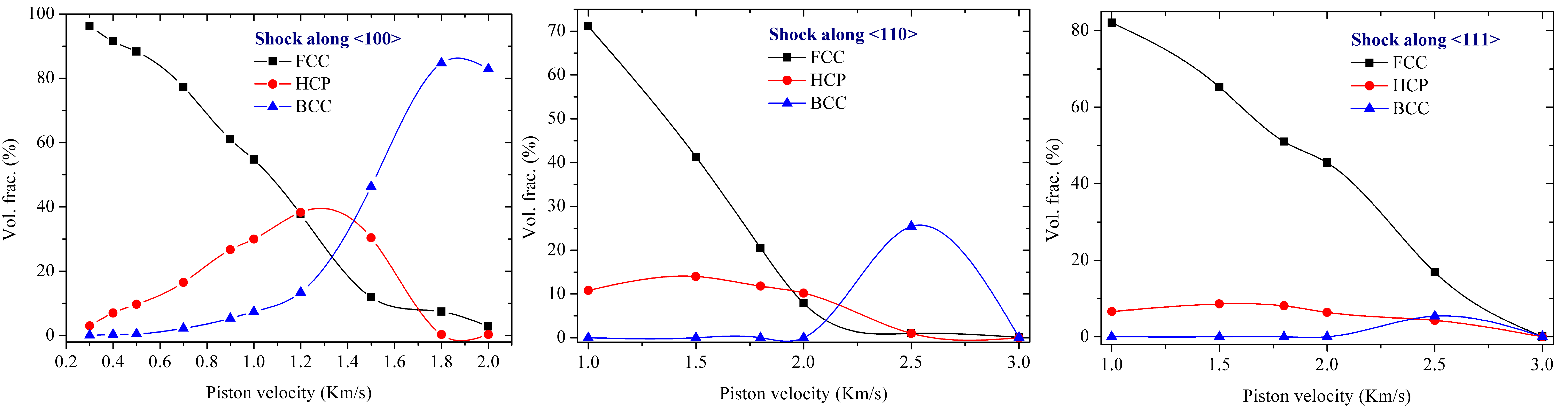}
  \caption{\label{vol-frac} Evolution statistics of different nucleated phases like BCC, HCP from initial undeformed FCC phase of copper at a range of piston velocity from 0.3 km/s to 3.0 km/s for three crystallographic direction of $<$100$>$, $<$110$>$ and $<$111$>$. The disordered phases (lattices with no definitive local crystal ordering as reported by adaptive-CNA\cite{Stukowski}) has not been presented here.}
\end{figure*}

The thickness of stacking faults steadily increases with increase in piston velocity thereby resulting in an increase in stacking fault density. Apart from development of stacking faults, FCC lattice structure is also observed to be transformed to HCP (hexagonal close pack) lattice structure along the fault planes through shear displacements. At a piston velocity of 1 km/s (ie. shock pressure of 56 GPa and temperature of 693.25 K), nearly the entire specimen is transformed into HCP through generation of very closely spaced stacking faults. The average distance between the consecutive faults becomes lesser with increase in piston velocity signifying corresponding enhancement of crystal plasticity; this has also been demonstrated earlier by Meyers et al.\cite{meyers1} by experimental investigations. At a critical value of shear stress, the length of the pinned dislocation segment becomes threshold and eventually bows out resulting in formation of twins\cite{johari}. Slip to twinning transformations are observed at piston velocities above 1 km/s. Infact at these high stress levels, twining becomes the preferred mode of deformation. Similar observations of onset, spreading and band formation of twining have been made through experimental investigations\cite{meyers1, cao}. Details about these different plasticity mechanisms for shock loaded Cu in different directions are presented by the authors in another journal submitted for publication.

Potential energy plot for different shock velocities of the Cu sample shock loaded in different direction is presented in Fig. \ref{pe} and a distinctive plateau region could be observed for the $<$100$>$ direction in between certain shock velocities. Increase in piston velocity results in higher excitation of the atoms which eventually leads to an increase in both kinetic and potential energy. The formation of the plateau region suggests that potential energy (radial distance from the reference coordinate) is not increasing significantly which indicates that some mechanisms are taking place which is able to dissipate the energy transmitted by the pistons. Even though this plateau region is observed for  $<$100$>$ direction, no such distinctive plateau region could be observed for  $<$110$>$ and  $<$111$>$ direction. Another feature which can be observed from Fig. \ref{pe} is that the shock equilibrated pressure for the $<$110$>$ and  $<$111$>$ directions are significantly higher than the $<$100$>$ direction. The reason for this can be explained as a result of high symmetry of the $<$100$>$ direction due to which the glissile movements are easily possible in that direction compared to the other two low symmetry directions. 

\begin{figure*}
\centering     
\subfigure []{\label{rdf100}\includegraphics[width=5.5cm]{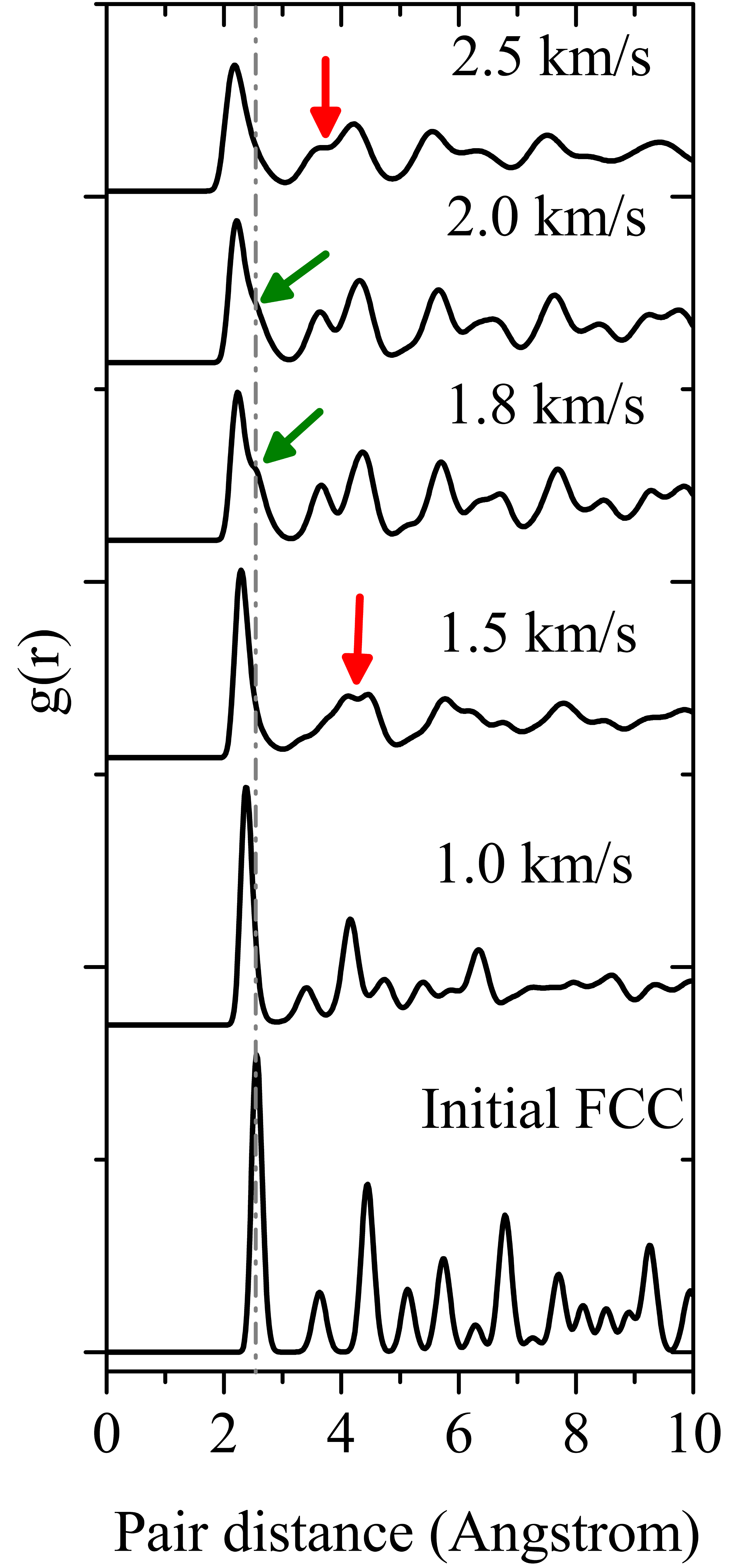}}
\subfigure []{\label{rdf110}\includegraphics[width=5.5cm]{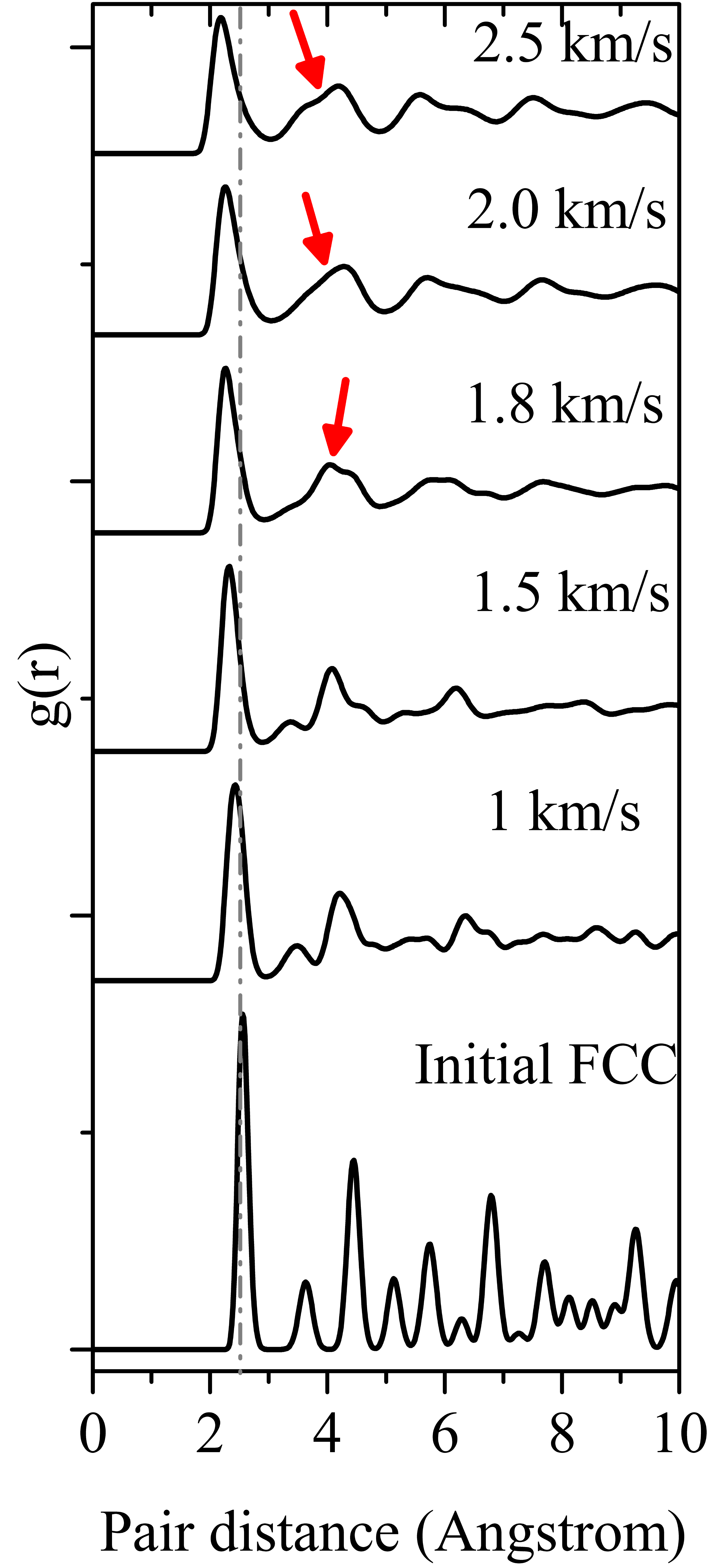}}
\subfigure []{\label{rdf111}\includegraphics[width=5.5cm]{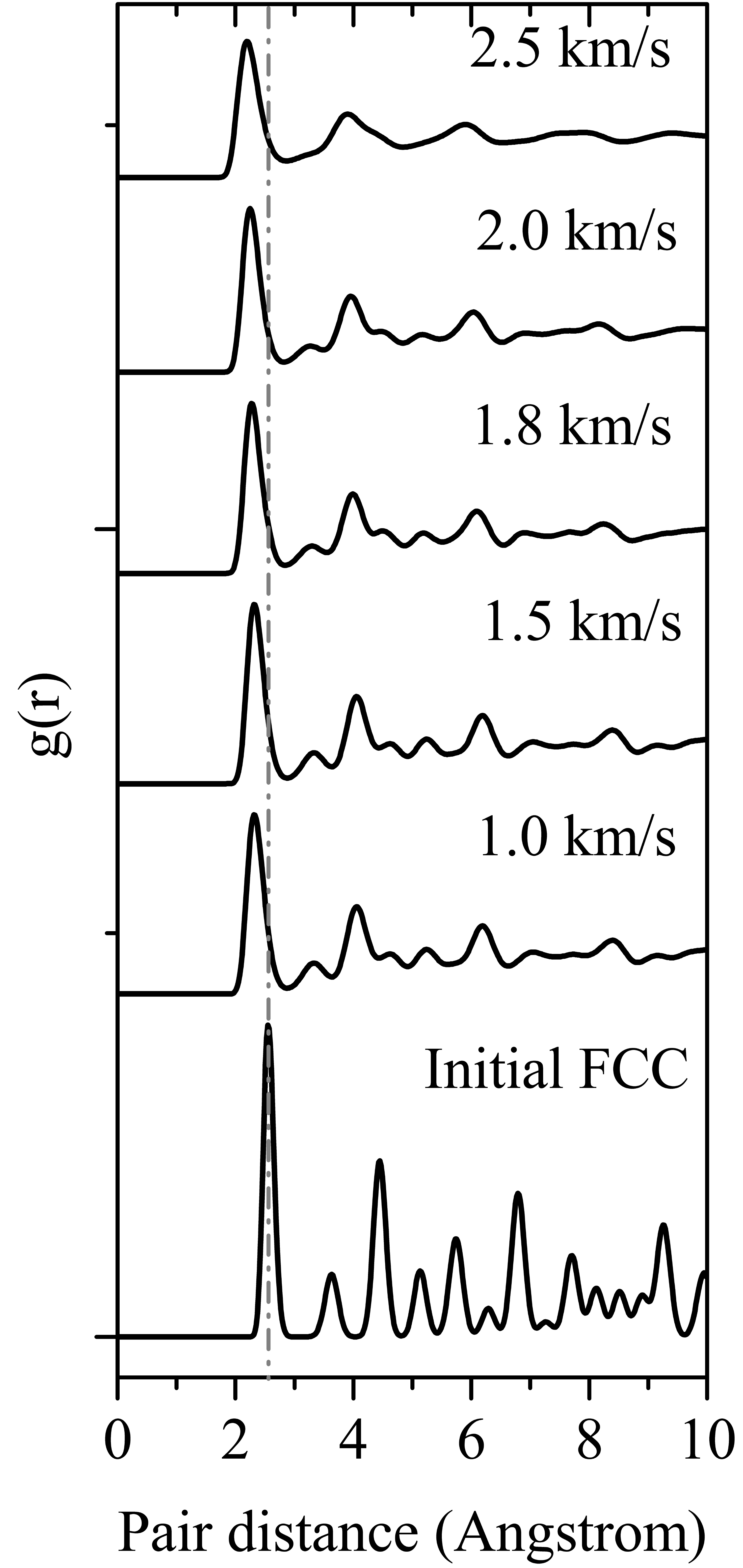}}
\caption{Calculated radial distribution function (RDF) of the deformed micro-structure shocked at a range of impact velocity of 1.0 km/s to 2.5 km/s, along the crystallographic direction (a)$<$100$>$, (b) $<$110$>$ and (c) $<$111$>$. In this stacked plots of RDFs, the gray colored dash-dot line represents the shift of the position of the first neighbour with respect to initial undeformed sample. Green arrows indicate the observance of the `kink' like shoulder inflamation region, whereas, red colored arrows point to the `wavy' nature of second neighbour position.}\label{microstructure-rdf}
\end{figure*}

Fig. \ref{vol-frac} shows volume fraction (expressed in percentage of the total volume) of different phases at different piston velocities for shock loading in different directions as obtained from adaptive CNA analysis. As discussed earlier, the adaptive CNA procedure even though gives a rough estimate of the phases present in a sample but cannot be specifically relied upon while determining the question of structural phase transition. Thereby there is a need to perform detailed analysis in order to clearly observe phase transition by using radial distribution functions (RDF) and x-ray diffraction analysis (XRD), which has been carried out later as part of this manuscript. The volume fraction figures (see Fig. \ref{vol-frac}) shows that for a piston velocities below 1 km/s the FCC volume fraction dominates other phases for all shock loading directions. Thereby there is almost no possibilities of structural phase transition at piston velocities at or below 1 km/s. As the piston velocities are increased, the volume fraction for BCC increases significantly in comparison to other phases only for shock loading along $<$100$>$ direction and thereby the probability of observing a phase transition increases. Thereby detailed studies are required to assess the possibility of phase transition, if any, at these intensities for $<$100$>$ direction. For shock loading along $<$110$>$ direction, an increase in volume fraction is observed for BCC (in comparison to other phases) only within a small band of piston velocities (2-2.5 km/s), where a detailed analysis is required. On the other hand, for shock loading along $<$111$>$ direction the volume fraction of BCC does not dominate the other phases significantly at any piston velocities, so obviously it can be expected that structural phase transition will not happen for shock loading in this direction for any piston velocities (till a piston velocity of 3.0 km/s as carried out in this research). 

Pair distribution function/Radial distribution function, g(r), calculates the probability of finding out two neighbors at a certain distance apart. Thereby RDF analysis can be used to determine the occurrence of a face-centered or a body-centered lattice structure for a particular material under a specific loading criterion. The RDF of the atoms (obtained from the trajectory of the shocked specimen upto 8 \AA) is calculated and averaged out over 5 ps time scale to obtain a clear picture regarding the temporal dependency as well as probability to obtain the 1st, 2nd and 3rd neighbors inside the specimen under shocked situation. Typically a body-centered phase has certain typical signatures such as presence of `shoulder region' in the descending part of the 1st coordination shell which indicates presence of two merged peaks. The peaks refer to the probabilities of atom neighbors near one another. In several situations a combination of thermal broadening of the peaks and the `shoulder region' results in formation of `kink' regions (or regions with two different slopes) in the descending part of the 1st coordination shell. These signatures of body-centered phase (typically the presence of `kinks') are observed from the RDF plot corresponding to $<$100$>$ shock loading direction for piston velocities of 1.8 to 2.0 km/s. The corresponding temperature and pressure ranges at these piston velocities are 1788-2286 K and 90-101 GPa respectively. 
Another typical signature of the possibility of presence of body-centered phase is merging of 2nd and 3rd coordination peaks resulting in a `wavy' nature of the RDF plot. This feature could be observed in for piston velocity of 1.5 km/s (in the $<$100$>$ shock loading direction) which does not show any clear peak for 2nd and 3rd coordination shell, rather the curve at this region of 2-4 \AA \, is `wavy' in nature. Similar `wavy' nature of the 2nd and 3rd coordination shell peaks could also be observed for 2.5 km/s piston velocity for the $<$100$>$ shock loading direction. A `wavy' nature of the 2nd and 3rd coordination shell is also observed for piston velocities of 1.8 to 2.5 km/s for the $<$110$>$ shock loading direction. No such signatures could be observed for the $<$111$>$ shock loading direction. Apart from these, general characteristics such as leftward shift of first coordination shell peak, thermal broadening and decrease in intensity of the peak as well as gradual loss of coordination typified by flattening nature of the higher coordination shell peaks with increase in piston velocities could be observed for shock loading in all directions. It should also be noted at this point that even though there are presence of certain signatures for body-centered phase from the RDF plots (which are reliable compared to adaptive CNA analysis) but these plots are not able to comment on whether the body-centered phase is cubic, tetrahedral, orthorhombic or some other lattice structures. It may also happen that due to presence of low volume fraction of the body-centered phase the effect/signature is not shown up in the RDF plots. Thereby the only way to definitively comment on observance of structural phase transition upon shock loading is through through XRD analysis. Virtual in-situ real-time XRD analysis (implemented in LAMMPS and validated for several materials with sufficient accuracy\cite{coleman}) of shock compressed samples is done for different shock intensities and for $<$100$>$ and $<$110$>$ orientations of shock loaded Cu sample to investigate in more detail the phenomena of shock induced structural phase transition. 

\begin{figure}[h!]
\centering
  \includegraphics[width=7.5cm]{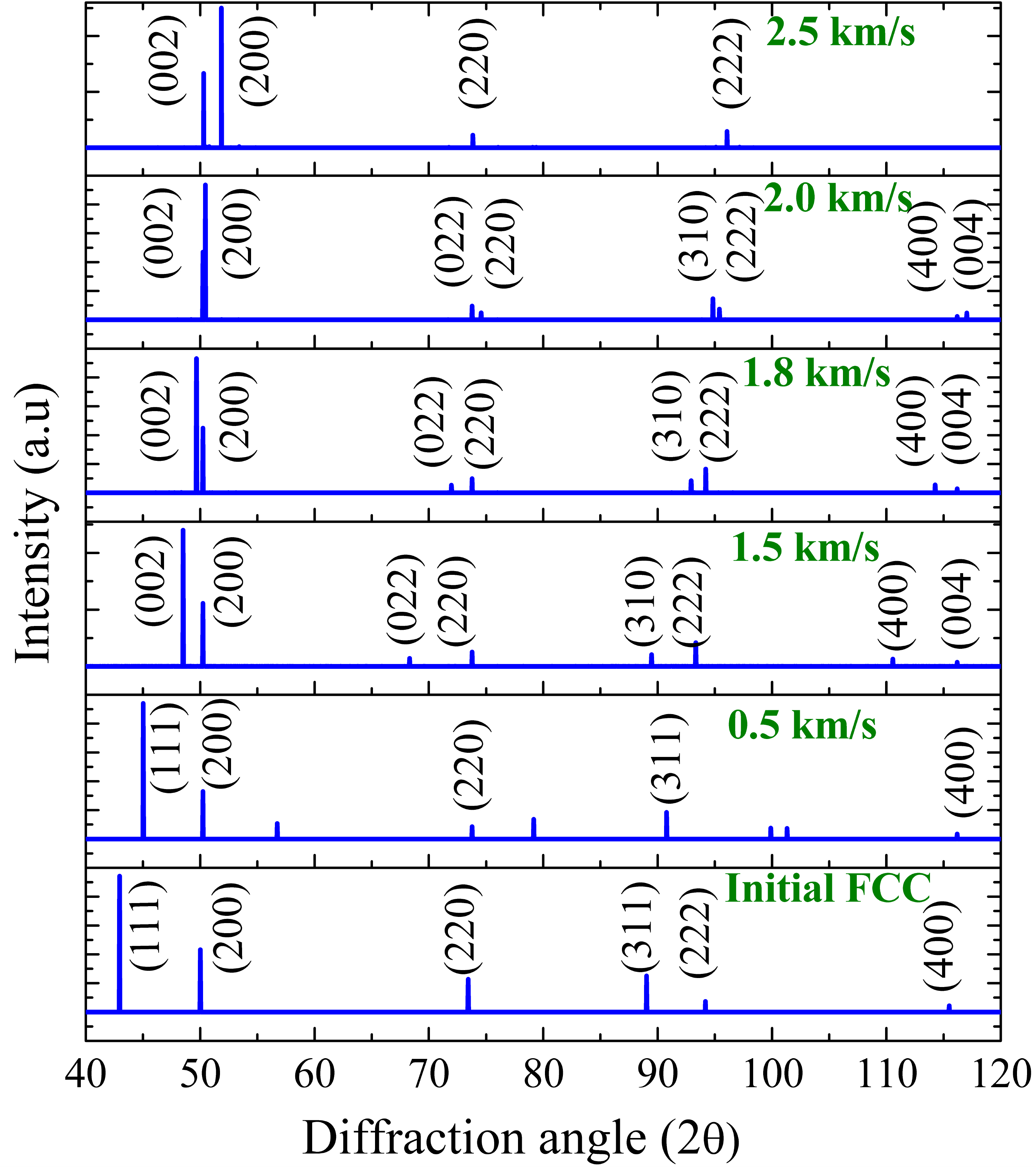}
  \caption{\label{xrd-100} X-ray diffraction pattern of initial undeformed FCC lattice of Cu and shock compressed (along $<$100$>$ crystallographic direction) Cu samples at piston velocity of 0.5, 1.5, 1.8, 2.0 and 2.5 km/s.}
\end{figure}

The Fig. \ref{xrd-100} shows XRD plot for final structure of shock loaded Cu in $<$100$>$ direction at different shock intensities. The primary peak corresponding to (111) plane was observed at a 2$\theta$ value of 42.92$^{\circ}$. The corresponding lattice parameter at the ambient temperature and pressure is a = 3.615 \AA \ (volume/atom = 11.68 \AA$^3$). The initial structure is an FCC structure since each peaks have either all odd or all even values of h, k, l\cite{cullity}. Peak indexing of the initial structure for Cu corresponds to a FCC crystal structure (which also match with JCPDS, copper file No. 04–0836). At a piston velocity of 0.5 km/s the peak corresponding to the (111) plane shifts to 45.05$^{\circ}$ which indicates deformation of the lattice. It should be noted that the vibration of atoms in a crystal leads to angle dependent effect on the diffracted peak intensities. Amongst the nine peaks observed in the figure, five peaks exactly match with FCC lattice with lattice constant a = 3.512 \AA. The presence of peaks 3, 5, 7, 8 indicates coexistence of another phase, possibly HCP. It is quite well known from literature that HCP phase originates on application of load to a FCC lattice of Cu, the identification of these peaks have not been rigorously done in this manuscript. It should be noted that the main intention of this paper is on identification of structural phase transformation to a body-centered phase. This distortion of the FCC structure (being demonstrated by shifts in the (111) plane) as well as increase in volume fraction of the HCP phase is observed for piston velocities below 1.5 km/s. 

At piston velocities of 1.5 km/s and above we observe even values of $(h+k+l)$ thereby demonstrating a body-centered structure\cite{cullity}. The Fig. \ref{xrd-100} shows that at a piston velocity of 1.5 km/s the (111) peak is no longer observed and the (200) peak (initially at 49.95$^{\circ}$ for the initial fcc structure) splits into 2 parts corresponding to (200) and (002) peaks demonstrating a typical characteristic of tetragonal structure formation. It should be pointed out that (002) plane spacing in a tetragonal structure differs from the other two planes ((200) and (020)); whereas all the three planes has same spacing for a cubic structure\cite{cullity}. The peaks (in 2$\theta$ angles) corresponding to (200) and (002) have been identified as 48.4$^{\circ}$ and 50.1$^{\circ}$. The `a' and `c' values are obtained as 3.63172 and 3.75286 \AA \, respectively with the `c/a' ratio of 1.033 (indicating a tetragonality of $\sim$3.3\%). Consistent with the splitting of (200) peak, the (220) peak in the initial FCC structure (at $73.48^{\circ}$) was observed to split into (022) and (220) (at 68.43$^{\circ}$ and 73.7$^{\circ}$) along with the splitting of (400) peak (initially at 115.49$^{\circ}$) to (004),(400) respectively (at 110.66$^{\circ}$ and 116.08$^{\circ}$) at the piston velocity of 1.5 km/s. At a piston velocity of 1.8 km/s the distances between the splitted peaks are observed to decrease - the angular distance (in 2$\theta$) between (200) and (002) was observed to decrease from 1.7$^{\circ}$ to 0.65$^{\circ}$ (49.73$^{\circ}$ and 50.38$^{\circ}$ respectively at 1.8 km/s). Similar decreases were also observed for the (220) and the (400) planes. The corresponding `a' and `c' values for piston velocity of 1.8 km/s are obtained as 3.6313 and 3.75012 \AA \, respectively. At a piston velocity of 2 km/s the two peaks ((002) and (200)) almost merge (49.95$^{\circ}$ and 50.38$^{\circ}$) and thereby the tetragonality is significantly reduced (`c/a' ratio being less than 1 \% with `a' as 3.63021 \AA \ and `c' as 3.63023 \AA).  The (022) and (220) peaks are observed at 73.78$^{\circ}$ and 74.57$^{\circ}$ whereas the (400) and (004) peaks are observed at 116.18$^{\circ}$ and 117.03$^{\circ}$. From the above discussion it can be confirmed that the crystal structure obtained on shock loading along $<$100$>$ direction (corresponding to piston velocities of 1.5 - 1.8 Km/s) is a body-centered tetragonal (BCT) structure.

At higher piston velocities (2.0 and 2.5 km/s) Fig. \ref{xrd-100} demonstrates the presence of (200) and (002) planes thereby demonstrating a BCT structure as explained before. It should be noted that even at these high temperatures and pressures (101.36 GPa, 2285.63 K and 134.28 GPa, 3931.81 K respectively) the crystal structure has not lost its periodicity and transformed into a amorphous/liquid state (which can be observed from the distinct peaks produced). Interestingly it can be observed that the intensities of the (200) and the (002) planes are reversed for 2.0 and 2.5 km/s piston velocity cases in comparison to what is observed for piston velocities 1.5 and 1.8 km/s. Typically peak intensities in XRD plots depends on structure factor, multiplicity factor, Lorentz polarizing factor, temperature and absorption factor. Under ambient temperature and pressure conditions, it is assumed that all atoms vibrate equally, which may not be the situation under high temperature and pressure. Usually with increase in temperature of the system, the Debye-Waller temperature factor increases which eventually leads to a reduction in the intensity and/or scattering amplitude. It may be possible that the atomic vibrations corresponding to the (002) plane is increased relative to that of the (200) plane at these higher piston velocities (2 - 2.5 km/s) which explains the reversal of intensities of these two planes in comparison to the intensities observed for these two planes for lower piston velocities (1.5 - 1.8 km/s). At 2.5 km/s piston velocity, the relative intensities of the (200) and (002) set of peaks are similar to that of 2.0 km/s piston velocity; however the distance between them in comparison to 2.0 km/s piston velocity has increased from 0.43$^{\circ}$ to 1.72$^{\circ}$. The corresponding peaks of (200) and (002) are observed at 50.32$^{\circ}$ and 52.04$^{\circ}$. Interestingly splitting of (220) peak is not observed as well as disappearance of (400) plane peaks are observed for 2.5 km/s piston velocity.
\begin{figure}[h!]
\centering
  \includegraphics[width=7.5cm]{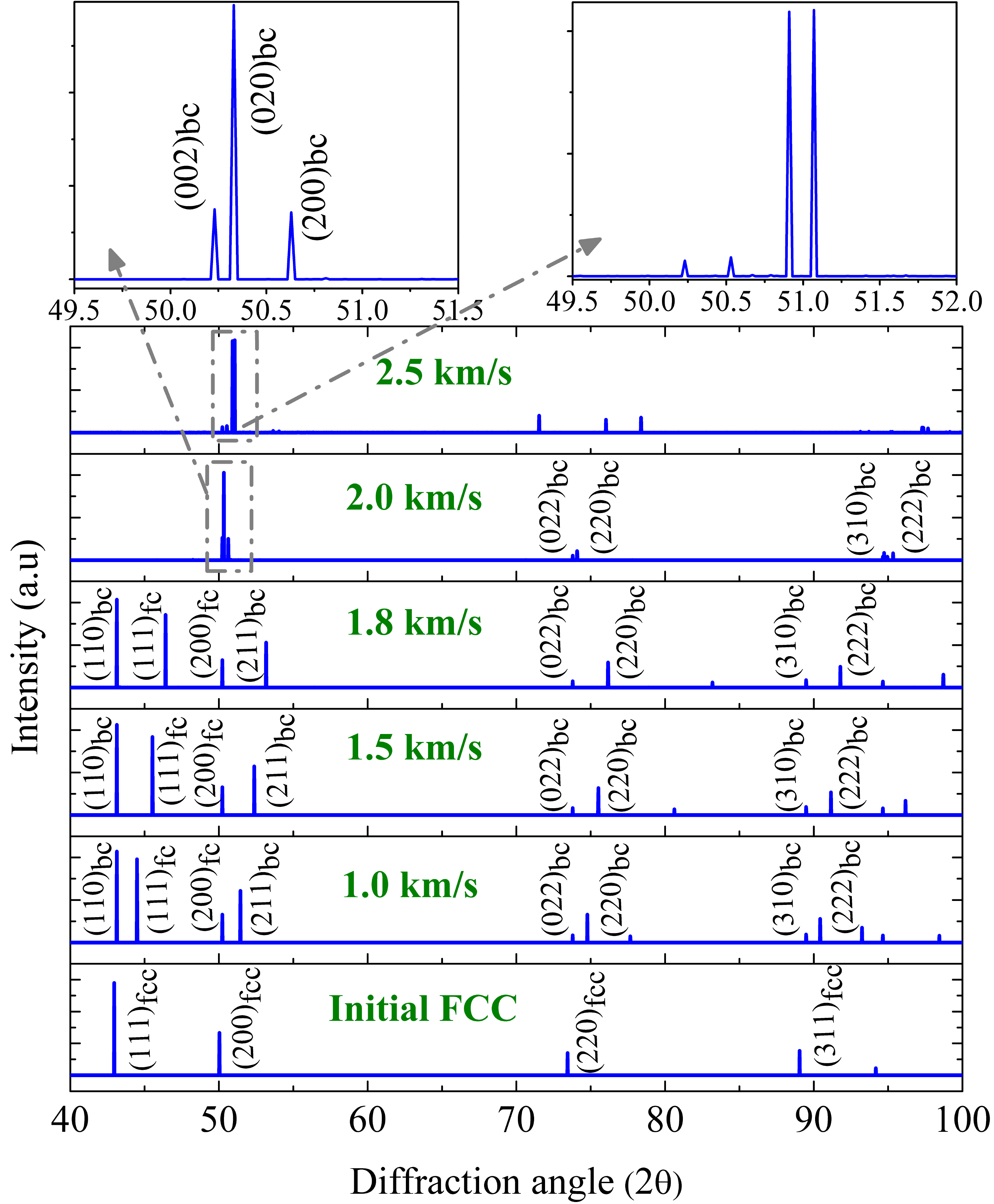}
  \caption{\label{xrd-110} X-ray diffraction pattern of initial undeformed FCC lattice of Cu and shock compressed (along $<$110$>$ crystallographic direction) Cu samples at piston velocity of 1.0, 1.5, 1.8, 2.0 and 2.5 km/s. The two boxes indicated by gray colored arrow represents the zoomed view of the splitting of the (002) plane for the impact velocity of 2.0 and 2.5 km/s. In the plots `bc' and `fc' indicates planes corresponding to body-centered and face-centered lattice respectively.}
\end{figure}
The results discussed above were for (100) direction of shock loading; discussion is being presented here for shock loading along (110) direction. 

The RDF analysis (refer Fig. \ref{rdf110}) shows `wavy' nature of the 2nd and the 3rd coordination shell indicating possibility of a body-centered phase for piston velocities above 1.8 km/s. However, it should be mentioned that RDF does not give good representation when there are combination of different phases in a system. Signatures of body-centered phases (even values of $(h+k+l)$) are observed from Fig. \ref{xrd-110} at a piston velocity of and above 1 km/s. However, the body-centered phase observed co-exists with FCC phase at piston velocities of 1 - 1.8 km/s. Since no splitting of peaks are observed, the body-centered phase observed can be classified as a body-centered cubic structure. At 2.0 km/s the (200) plane is observed to split into (002), (020) and (200) thereby demonstrating formation of an orthorhombic phase\cite{cullity}, which was earlier postulated by Hirth et al.\cite{hirth}. The peaks of body-centered orthorhombic (BCO) phase of copper at (002), (020) and (200) plane are observed at 50.23, 50.33 and 50.63$^{\circ}$ respectively. However the higher order planes such as (220) only splits into (220) and (022) which is typically observed for tetragonal structures. At 2.5 km/s, four peaks can be observed near to a 2$\theta$ angle of $\sim$50$^{\circ}$ and the structural characteristics could not be identified with ease. It should be noted that orthorhombic structure is not demonstrated at this piston velocity and it should also be pointed out that amorphous/melting was also not observed at this piston velocity. Since no signatures of `kink' formation or `wavy' nature could be obtained from Fig. \ref{rdf111}, the detailed XRD analysis was not performed for the $<$111$>$ direction of shock loading. 

\section{CONCLUSION}
A new phase of Cu has been identified through this research in which Cu shock loaded along $<$100$>$ direction with piston velocities of 1.5 to 2.5 km/s results in formation of a BCT phase. BCC phase of Cu has been observed to co-exist with deformed FCC structure when Cu is shock loaded along $<$110$>$ direction for piston velocities of 1-1.8 km/s. For shock loading along $<110>$ direction, at 2.0 km/s piston velocity, a BCO phase of Cu has been observed (which has been postulated earlier by Hirth et al.\cite{hirth}). Shock loading along $<111>$ direction did not produce any evidence of body-centered phase at different piston velocities. The numerical simulations demonstrating novel phases of Cu are yet to be validated through experimental observations in which real time XRD analysis is to be coupled with shock loading test set-ups. It should also be mentioned that where these novel phases of Cu were obtained, the plasticity mechanisms corresponding to dislocation and twins are not observed. 

\section{ACKNOWLEDGMENTS}
A.N. gratefully acknowledge Dr. Abhijit Ghosh, Sk. Md. Hasan and Anish Karmakar for useful discussions.

\end{document}